\documentclass[aps,prl,twocolumn,showpacs,superscriptaddress]{revtex4-1}

\usepackage{amsfonts}
\usepackage{amsmath}
\usepackage{amssymb}
\usepackage{graphicx}
\usepackage{bm}

\begin{document}



\title{ Interplane resistivity of underdoped single crystals (Ba$_{1-x}$K$_x$)Fe$_2$As$_2$, $0 \leq x<0.34$}


\author{M.~A.~Tanatar}
\email[Corresponding author: ]{tanatar@ameslab.gov}
\affiliation{Ames Laboratory, Ames, Iowa 50011, USA}
\affiliation{Department of Physics and Astronomy, Iowa State University, Ames, Iowa 50011, USA }

\author{W. E. Straszheim}
\affiliation{Ames Laboratory, Ames, Iowa 50011, USA}

\author{Hyunsoo~Kim}

\affiliation{Ames Laboratory, Ames, Iowa 50011, USA}
\affiliation{Department of Physics and Astronomy, Iowa State University, Ames, Iowa 50011, USA }

\author{J.~Murphy}

\affiliation{Ames Laboratory, Ames, Iowa 50011, USA}
\affiliation{Department of Physics and Astronomy, Iowa State University, Ames, Iowa 50011, USA }

\author{N.~Spyrison}

\affiliation{Ames Laboratory, Ames, Iowa 50011, USA}
\affiliation{Department of Physics and Astronomy, Iowa State University, Ames, Iowa 50011, USA }

\author{E.~C.~Blomberg}

\affiliation{Ames Laboratory, Ames, Iowa 50011, USA}
\affiliation{Department of Physics and Astronomy, Iowa State University, Ames, Iowa 50011, USA }

\author{K.~Cho}

\affiliation{Ames Laboratory, Ames, Iowa 50011, USA}
\affiliation{Department of Physics and Astronomy, Iowa State University, Ames, Iowa 50011, USA }

\author{J.-Ph.~Reid}
\affiliation{D\'epartement de physique \& RQMP, Universit\'e de Sherbrooke, Sherbrooke, Qu\'ebec, Canada J1K 2R1}

\author{ Bing Shen }
\affiliation{ Institute of Physics, Chinese Academy of Sciences, Beijing 100190, P. R. China }

\author{Louis Taillefer}
\affiliation{D\'epartement de physique \& RQMP, Universit\'e de Sherbrooke, Sherbrooke, Qu\'ebec, Canada J1K 2R1}
\affiliation{Canadian Institute for Advanced Research, Toronto, Ontario, Canada M5G 1Z8}

\author{Hai-Hu Wen}
\affiliation{ Institute of Physics, Chinese Academy of Sciences, Beijing 100190, P. R. China }
\affiliation{ National Laboratory of Solid State Microstructures and Department of Physics, Nanjing University, Nanjing 210093, P. R. China }
\affiliation{Canadian Institute for Advanced Research, Toronto, Ontario, Canada M5G 1Z8}

\author{R.~Prozorov}
\affiliation{Ames Laboratory, Ames, Iowa 50011, USA}
\affiliation{Department of Physics and Astronomy, Iowa State University, Ames, Iowa 50011, USA }

\date{11 April 2014}


\begin{abstract}

Temperature-dependent inter-plane resistivity, $\rho _c(T)$, was  measured in hole-doped iron-arsenide superconductor
(Ba$_{1-x}$K$_x$)Fe$_2$As$_2$ over a doping range from parent compound to optimal doping $T_c\approx 38~K$, $0\leq x \leq 0.34$. Measurements were undertaken on high-quality single crystals grown from FeAs flux. The coupled magnetic/structural transition at $T_{SM}$ leads to clear accelerated decrease of $\rho_c(T)$ on cooling in samples with $T_c <$26~K ($x <0.25$). This decrease in hole-doped material is in notable contrast to an increase in $\rho_c(T)$ in the electron-doped Ba(Fe$_{1-x}$Co$_x$)Fe $_2$As$_2$ and iso-electron substituted BaFe$_2$(As$_{1-x}$P$_x$)$_2$. The $T_{SM}$ decreases very sharply with doping, dropping from $T_s$=71~K to zero on increase of $T_c$ from approximately 25 to 27~K. The $\rho_c(T)$ becomes $T$-linear close to optimal doping.
The broad crossover maximum in $\rho_c(T)$, found in the parent BaFe$_2$As$_2$ at around $T_{max} \sim$200~K, shifts to higher temperature $\sim$250~K with doping $x$=0.34. The maximum shows clear correlation with the broad crossover feature found in the temperature-dependent in-plane resistivity $\rho_a(T)$. The doping evolution of $T_{max}$ in (Ba$_{1-x}$K$_x$)Fe$_2$As$_2$ is in notable  contrast with both rapid suppression of $T_{max}$ found in Ba(Fe$_{1-x}TM_x$)$_2$As$_2$ ($TM$=Co,Rh,Ni,Pd) and its rapid increase BaFe$_2$(As$_{1-x}$P$_x$)$_2$.
This observation suggest that pseudogap features are much stronger in hole-doped than in electron-doped iron-based superconductors, revealing significant electron-hole doping asymmetry similar to the cuprates.

\end{abstract}



\maketitle



\section{Introduction}

Superconductivity in hole-doped (Ba$_{1-x}$K$_x$)Fe$_2$As$_2$ \cite{Rotter} (BaK122, in the following) was found soon after the discovery of superconductivity with high critical temperatures in oxypnictide FeAs-based materials \cite{Hosono}.
Intensive studies of the doping phase diagram were undertaken on high quality polycrystalline materials using neutron scattering, magnetization, heat capacity and pressure-dependent measurements \cite{Rotterheatcap,Avci1,Avci2,Budko}. They revealed that similar to electron-doping in Ba(Fe$_{1-x}TM_x$)$_2$As$_2$ ($TM$=Co,Rh,Ni,Pd, Ba$TM$122 in the following) \cite{CB} and isoelectron substitution in BaFe$_2$(As$_{1-x}$P$_x$)$_2$ \cite{Kasahara} (BaP122 in the following), maximum $T_c$ is observed close to a point where magnetism vanishes, suggesting possible existence of the quantum critical point (QCP) in the phase diagram \cite{NDL,PQCP} and magnetically mediated pairing \cite{Lonzarich,Norman}.

A hallmark of this scenario is systematic evolution of the temperature-dependent resistivity, $\rho (T)$, over the phase diagram. Typically the $\rho (T)$ is close to $T$-linear at optimum doping and to $T^2$ in the overdoped regime \cite{NDL,Tailleferreview}, while at inter-mediate compositions it can be represented by either a power-law function $\rho (T)= \rho_0 +\rho_n T^{n}$, or as a sum of linear and quadratic terms, $\rho (T)= \rho_0 +\rho_1 T+ \rho_2 T^2$. Interestingly, the magnitude of the $T$-linear contribution to resistivity correlates with the superconducting $T_c$, providing important link between anomalous scattering and pairing \cite{Tailleferreview}.
This doping-dependent $\rho (T)$ and a $T$-linear dependence at optimal doping are indeed observed in both in-plane, $\rho_a (T)$, and inter-plane, $\rho_c(T)$, resistivity of  BaP122 \cite{Kasahara,BaPinterplane}, revealing clear signatures of quantum critical point both in normal \cite{PQCP} and superconducting \cite{ScienceHashimoto} states. 

The situation is clearly more complicated in both electron-doped BaCo122 and hole-doped BaK122. In both cases the doping-dependent $T_N(x)$ was found to be non-monotonic with reentrance of the tetragonal phase  \cite{Nandi,Avci1}, suggesting no true existence of quantum critical point in the phase diagram. 
Despite this, the in-plane transport in BaCo122 reveals systematic evolution from $T$-linear to $T^2$ on going from optimal doping to overdoped compositions, as expected for QCP scenario, however, the inter-plane resistivity, $\rho_c(T)$, reveals $T$-linear dependence only in narrow range above $T_c$, terminated at high temperatures by a broad crossover maximum at $T_{max}$ \cite{anisotropy,pseudogap}. Similar maximum is observed in $\rho_c(T)$ of all transition metal electron-doped Ba$TM$122 \cite{pseudogap2}. By correlation with $T$-linear increase of magnetic susceptibility and NMR Knight shift, we related the maximum at $T_{max}$ with pseudogap \cite{pseudogap}, the existence of which was first suggested by NMR studies in electron-doped BaCo122 \cite{pseudogapNMR1,pseudogapNMR2}. The pseudogap region extends from parent compound to far beyond the end of the superconducting dome in the doping phase diagram for electron-doped BaCo122 \cite{pseudogap,pseudogap2}. The existence of pseudogap in iron based superconductors was later confirmed with spectroscopic \cite{Moon1,Moon2} and ARPES \cite{Shimojima} techniques. 

Pseudogap is one of the dominant puzzling features in the phase diagram of the hole-doped cuprates \cite{TimuskStatt}. On the other hand, its effect on the properties of electron-doped high-$T_c$ cuprates is not so pronounced \cite{RLGreen}. It was suggested that $T$-linear in-plane resistivity in the cuprates is determined by the quantum critical point of the pseudogap phase \cite{pseudogapQCP}, and is linked with the competing nematic ordering \cite{pseudogapnematic}. These discussions strongly influence studies of the QCP scenario, nematicity and of the pseudogap in iron based superconductors \cite{RafaelNP}. Previously, we have shown that electronic nematicity of 122 family of iron based superconductors is strongly suppressed on the hole - doped side of the phase diagram and even changes sign \cite{BaKdetwinning}. Therefore, it of prime interest whether the electron-hole doping asymmetry is also characteristic for the pseudogap features and QCP in iron pnictides. With this motivation in mind here we report a systematic study of the inter-plane (c-axis) transport in hole-doped iron based superconductor (Ba$_{1-x}$K$_x$)Fe$_2$As$_2$.

Previous studies of the doping evolution of the temperature-dependent in-plane resistivity in BaK122 \cite{Wencrystals,BingShen} found that when data are analyzed using a power-law function, $\rho (T)= \rho_0 +\rho_n T^{n}$, the exponent $n$ of the fit monotonically decreases on approaching optimal doping from the under-doped side, however, it always remains higher than one. Analysis of the frequency-dependent optical conductivity \cite{Holms} of optimally doped BaK122 suggested that in fact $T$-linear term in resistivity is masked by the existence of two Drude contributions to conductivity, only one of which is $T$-linear. Similar multi-component analysis of conductivity was suggested by Golubov {\it et al.} \cite{Zverev} to explain resistivity crossover at around 200~K. The authors considered the model in which two contributions to conductivity have very different $\rho (T)$. The one with low residual resistivity and strong $T$-dependence is dominating low-temperature part 
 of $\rho(T)$, while the one with high residual resistivity and weak $T$-dependence becomes dominant at high temperatures. Alternatively the $\rho_a(T)$ of BaK122 was fitted by Gasparov {\it et al.} \cite{Gasparov} using $\rho(T)=\rho_0+\rho_nT^n+\rho_e exp(-T_0/T)$, with the third term arising from phonon-assisted scattering between two Fermi-surface sheets.

Pressure studies of the underdoped BaK122 crystals by Hassinger {\it et al.} \cite{Elena} found an anomaly due to intervening new phase in the doping range close to compositional edge of the magnetism, with the anomaly in in-plane transport of the crystals with $T_{SM} \sim$95~K. An anomaly in similar doping range was found at ambient pressure in sign-reversal of in-plane resistivity anisotropy of BaK122 \cite{BaKdetwinning} and in high-quality polycrystalline samples of another hole doped composition, BaNa122 \cite{BaNa}.

As can be seen, there is no systematic picture of doping evolution of the transport properties in hole-doped BaK122. Additional problem comes from the fact that properties of the samples of BaK122 grown using different fluxes are different. Sn-grown parent Ba122 shows quite significant suppression of $T_{SM}$ down to ~90~K \cite{NiNiSn}, compared to approximately 135~K \cite{CB} in FeAs flux crystals or polycrystalline materials \cite{Rotter,Avci1}. This strong suppression is ascribed to incorporation of Sn at sub-percent level \cite{Sn-private}. That is why the goal of this study is to characterize the doping evolution of the temperature-dependent resistivity in high-quality single crystals of BaK122 grown from FeAs flux.

In this article we report systematic study of inter-plane resistivity of single crystals of BaK122, grown using FeAs flux technique. Our main findings may be summarized as follows. (1) The pseudogap crossover maximum observed in $\rho_c (T)$ at $T_{max}$ moderately shifts to higher temperatures with $x$ in BaK122, significantly slower than it does in iso-electron substituted BaP122 \cite{BaPinterplane} and with opposite trend to electron-doped Ba$TM$122 \cite{pseudogap,pseudogap2}. (2) The cross-over maximum correlates well with a slope-change feature in temperature-dependent in-plane resistivity, suggesting its relation to carrier activation. (3) A range of $T$-linear dependence is observed in inter-plane resistivity of close to optimal doping BaK122, in contrast to the slightly super-linear dependence with $n$=1.1 of the in-plane transport \cite{BingShen}. (4) The anomalies found in the pressure studies of the underdoped samples are not found reproducibly in the doping study, suggesting a difference in hole-doping and pressure-tuned phase diagrams.

\section{Experimental}

\subsection{Sample preparation}

Single crystals of BaK122 were grown using high temperature FeAs flux technique \cite{Wencrystals}.
The volatility of K during growth leads to distribution of potassium content, with the inner parts of the crystals frequently having $T_c$ differing by 1 to 3 K from the surface parts.
Because of this distribution, as a first step in sample preparation for our study, we cleaved thin slabs from the inner part of the crystals, typically of 20 $\mu$m thickness. The slabs had two clean and shiny cleavage surfaces. The samples were cleaved from these slabs with sides along (100) directions using razor blade. They typically had dimensions of 0.5$\times$0.5$\times$0.02 mm$^3$ size ($a \times b \times c$)

We used two protocols for sample characterization for inter-plane resistivity measurements. All samples were prescreened using dipper version of the TDR technique \cite{NaFeAs1,NaFeAs2}, using the sharpness of the superconducting transition as a measure of constant dopant concentration in each particular piece. These measurements also allowed us to exclude possible inclusions with lower $T_c$.
After this pre-screening, samples with the most sharp transitions were characterized by magneto-optical technique to look for possible inhomogeneity, as described in detail in Ref.~\onlinecite{vortex,MO,SUST}, and then their chemical composition was determined using wavelength dispersive x-ray spectroscopy (WDS) in JEOL JXA-8200 electron microprobe. The composition was measured for 12 points per single crystal and averaged.
We refer to this group of samples as group A in the following.

Inter-plane resistivity was measured on all crystals studied in WDS to determine resistive $T_c$ and structural transition temperature, $T_{SM}$ as a function of composition $x$. For this purpose top and bottom surfaces of the samples were covered with Sn solder \cite{SUST,patent} and 50 $\mu$m silver wires were attached to enable measurements in four-probe configuration.
Soldering produced contacts with resistance typically in the 10 $\mu \Omega$ range. Inter-plane resistivity was measured using a two-probe technique with currents in 1 to 10 mA range (depending on sample resistance which is typically 1 m$\Omega$), relying on the negligibly small contact resistance. Four-probe scheme was used down to the sample to measure series connected sample, $R_s$, and contact, $R_c$ resistance. Taking into account that $R_s \gg R_c$, contact resistance represents a minor correction of the order of 1 to 5\%. This can be directly seen for our samples for temperatures below the superconducting $T_c$, where $R_s =$0 and the measured resistance represents $R_c$ \cite{anisotropy,SUST,vortex}.
The details of the measurement procedure can be found in Refs.~\onlinecite{anisotropy,anisotropypure,pseudogap}.

The drawback of the measurement on samples with $c \ll a$ is that any inhomogeneity in the contact resistance or internal sample connectivity admixes the in-plane component due to redistribution of the current.
This requires measurements on a bigger array of samples, beyond our possibility of WDS measurements. To check for reproducibility, we performed $\rho_c$ measurements on samples which had same dipper TDR $T_c$ as samples of group A. We refer to these samples as group B. We performed measurements of $\rho_c$ on at least 5 samples of each batch with the same dipper TDR $T_c$, at least one of the samples was measured in WDS to determine composition. In all cases we obtained qualitatively similar temperature dependencies of the electrical resistivity, as represented by the ratio of resistivities at room and low temperatures, $\rho _c (0)/\rho _c (300)$. The resistivity value, however, showed a notable scatter and at room temperature, $\rho_c(300K)$, was typically in the range 1000 to 2000 $\mu \Omega$~cm.

\begin{figure}[tbh]%
\centering
\includegraphics[width=8.7cm]{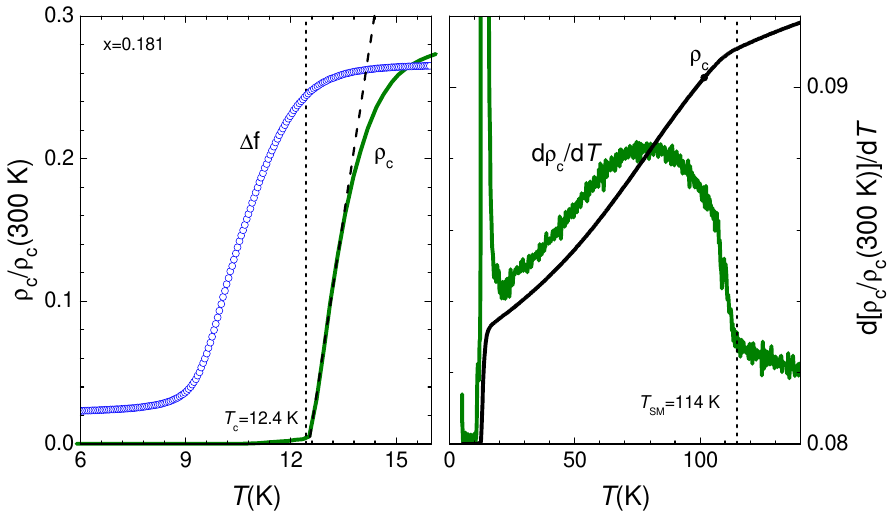}%
\caption{(Color online) Left panel. Temperature dependent resistivity and frequency shift in TDR measurements (shown in arbitrary units) in the superconducting transition range, used to determine superconducting $T_c$ of samples of group A. Right panel zooms the $\rho_c(T)$ in the area of the structural transition (arbitrary scale), showing criterion used to determine $T_{SM}$.
}%
\label{TcrhoTDR}%
\end{figure}

Because $\rho_c(T)$ measurements are made in the two-probe mode, resistivity value below $T_c$ is always finite and represents contact resistance. Therefore the superconductive transition temperature was determined as an offset point of the sharp part of the resistive transition, as shown in Fig.~\ref{TcrhoTDR}. For reference we show in the same graph the temperature-dependent TDR frequency shift of the same sample before contact application. We find that the offset point of the superconducting transition in $\rho_c(T)$ measurements corresponds well to the onset point in $\Delta f (T)$. In the right panel of Fig.~\ref{TcrhoTDR} we show temperature dependent inter-plane resistivity of the same sample in the temperature range of the structural/magnetic transition. We also show temperature dependence of resistivity derivative, $d[\rho_c(T)/\rho_c(300K)]/dT$. Of note that contrary to the in-plane resistivity, structural/magnetic transition leads to a resistivity decrease and onset 
 of increase of the resistivity derivative. We define $T_{SM}$ as an onset point of rapid rise in $d[\rho_c(T)/\rho(300K)]/dT$. Note that resistivity derivative has singular feature at the transition, contrary to split double-feature structure observed in BaCo122 \cite{CB}. This is consistent with the coincident tetragonal-to-orthorhombic and antiferromagnetic transitions as found in neutron scattering experiments \cite{Avci1}.

\section{Results}

\subsection{ WDS composition analysis}

\begin{figure}[tbh]%
\centering
\includegraphics[width=8.5cm]{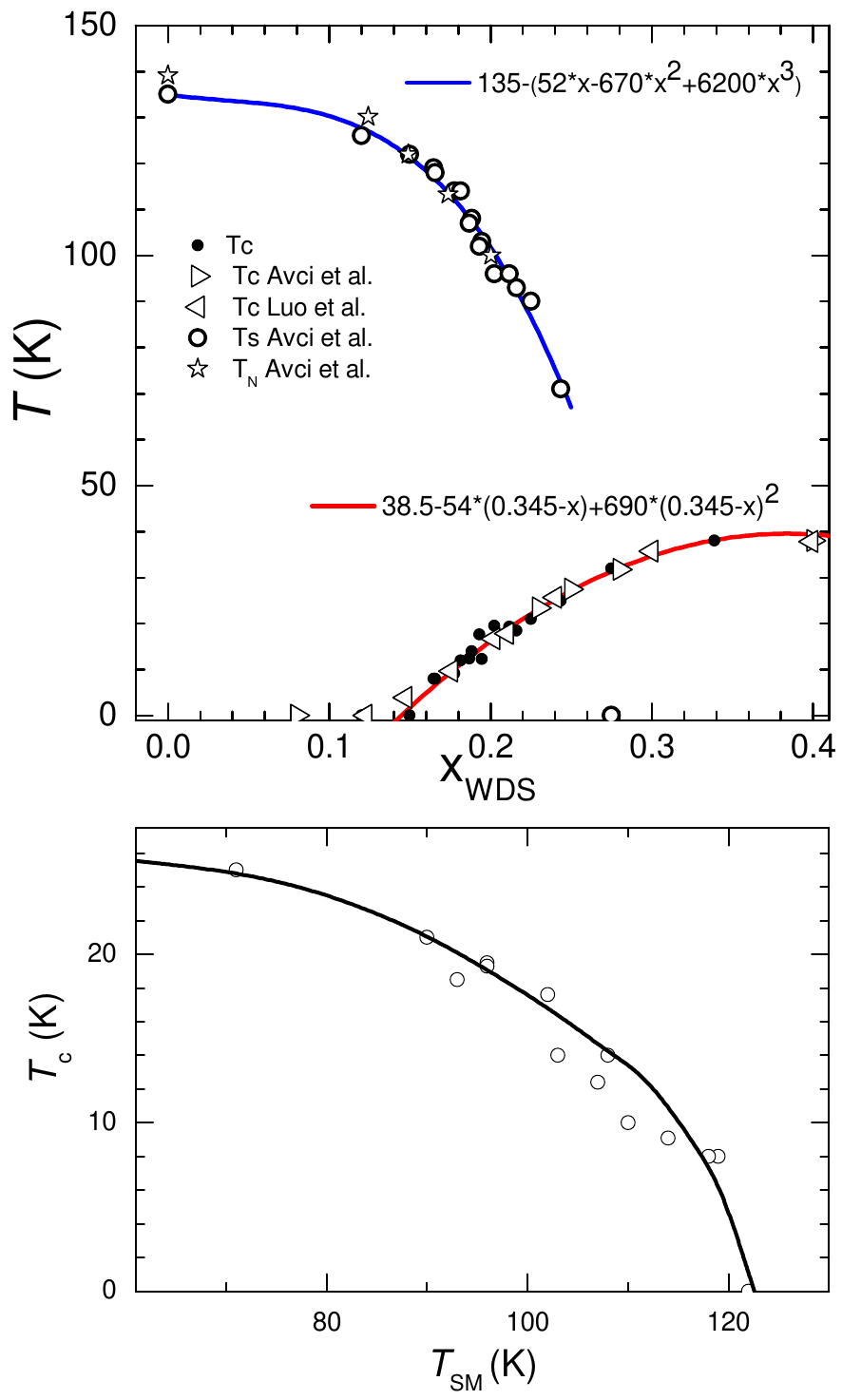}%
\caption{(Color online) Top panel. The structural/magnetic transition temperature, $T_{SM}$, (open circles) and the superconducting $T_c$ (solid circles) as a function of $x$ determined in WDS measurements on the same crystals of group A. The definition of $T_c$ and $T_{SM}$ is shown in Fig.~\ref{TcrhoTDR}. For reference we show $T_c (x)$ (right open triangles) as determined from magnetization measurements and $T_{SM}(x)$ (open stars) as determined from neutron scattering measurements on high quality polycrystalline samples,  \cite{Avci1,Avci2} and $T_c(x)$ as determined from resistivity measurements (open left triangles) on single crystals grown out of FeAs flux \cite{Wencrystals}.
Lines show fit through the data for $T_{SM}(x)$ (blue) and $T_c(x)$ (red) curves. 
Bottom panel shows $T_c(T_N)$ dependence, as determined in our measurements on the same crystals. This dependence is monotonic, suggesting no doping anomaly in either $T_c$ or $T_N$. }%
\label{WDS}%
\end{figure}

In Fig.~\ref{WDS} we show evolution of the temperatures of structural/magnetic and superconducting transitions in the crystals of group A. For reference we show data obtained on high quality polycrystalline materials, and in the previous study on single crystals \cite{Wencrystals}. The three studies are in reasonable agreement with the minor difference being in determinations at the very edge of the superconducting dome. Our study suggests that superconductivity sets in at $x$=0.15, which is somewhat higher than the value found by Avci {\it et al.} \cite{Avci1}. 

The doping evolution of the structural transition temperature $T_{SM}$ is in reasonable agreement with neutron scattering data of Avci \cite{Avci1}. The fit of $T_{SM}(x)$ requires third order polynomial and is not very precise. Of note, despite quite small steps in $T_c$ of the samples we have not found samples with structural transition temperatures below 71~K, which may suggest very sharp termination of $T_{SM}(x)$. 

The dependence of zero-resistivity $T_c$ can be well fit over the range studied using a parabolic function, $T_c$=38.5-54*(0.345-x)+690*(0.345-x)$^2$, as shown with solid line. This parabolic dependence is similar to the parabolic dependence found in the cuprates \cite{Tallon}. Contrary to the cuprates, though, this dependence is very asymmetric in BaK122 \cite{Rotter,Avci1}, it obviously fails in the overdoped regime. 

The samples shown in the top panel of Fig.~\ref{WDS} have monotonic relation between $T_c$ and $T_{SM}$, similar to the behavior found in previous studies on BaK122 polycrystals \cite{Rotter,Avci1,Avci2} and for other doping types \cite{Paglione,Johnston,Stewart,CB}. This dependence provides intrinsic check for sample quality and is shown in the bottom panel of Fig.~\ref{WDS} with solid dots. This monotonic dependence is in striking contrast with measurements under pressure by Hassinger {\it et. al.}  \cite{Elena}, finding a competing phase reducing $T_c$ from its doping trend, which thus should lead to an anomaly in $T_c(T_N)$. This observation is suggestive that doping and pressure phase diagrams are not quite equivalent in BaK122 system, 
This unusual difference of the two tuning parameters is not found in electron-doped BaCo122 \cite{Colombier}, in which pressure and doping lead to similar $T_c$ evolution. It is also different from the behavior in iso-electron-substituted BaRu122 \cite{Stella}. On the contrary, difference between doping and pressure tuning was found in thin-films of indirectly electron doped BaLa122 \cite{Hosono2}, in which $T_c$ monotonically increases with pressure in both under-doped and over-doped regimes.

\subsection{ Maximum of inter-plane resistivity }

\begin{figure}[tbh]%
\centering
\includegraphics[width=8.7cm]{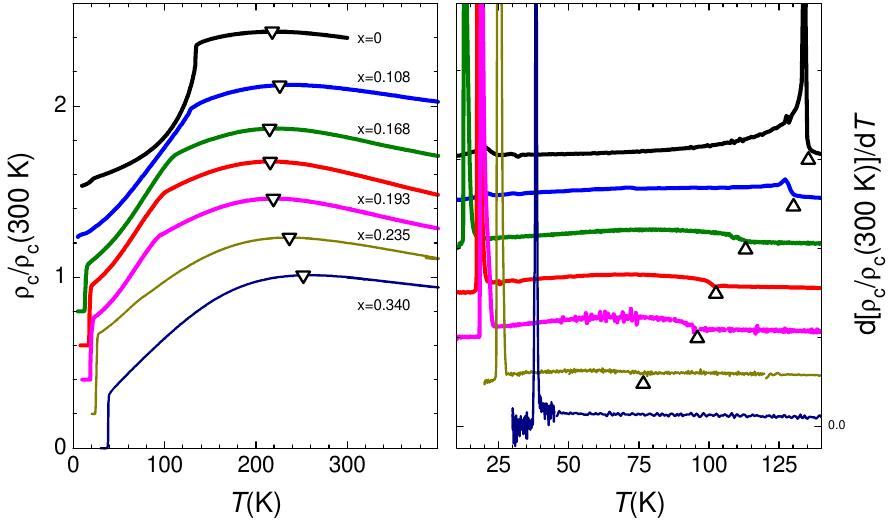}%
\caption{(Color online) Left panel shows doping evolution of the temperature-dependent inter-plane resistivity, $\rho_c(T)$, normalized to room temperature values, $\rho _c (300K)$. The curves are offset to avoid overlapping. Down triangles show position of the $\rho _c (T)$ maximum at $T_{max}$. Right panel shows temperature dependent resistivity derivative, with up-triangles showing position of the $T_{SM}$. }%
\label{rhocdoping}%
\end{figure}

In Fig.~\ref{rhocdoping} we show doping evolution of the temperature-dependent inter-plane resistivity in BaK122. In addition to features due to magnetic/structural transition at $T_{SM}$ and superconductivity at $T_c$, discussed above, $\rho_c(T)$ shows a clear maximum, observed in parent Ba122 at $T_{max}\approx$200~K. Because of the broad crossover character and possible influence on maximum position of the admixture of $\rho_a(T)$ \cite{anisotropy}, maximum is defined with rather large error bars of about $\pm$20~K. The doping up to $x$=0.235, on the edge of the orthorhombic/antiferromagnetic domain in the phase diagram, does not change the position of the maximum within the error bars, doping to $x$=0.34, close to optimal doping, slightly shifts $T_{max}$ to higher temperatures.

\section{Discussion}

\subsection{Structural/magnetic ordering and inter-plane resistivity}

\begin{figure}[tbh]%
\centering
\includegraphics[width=8.5cm]{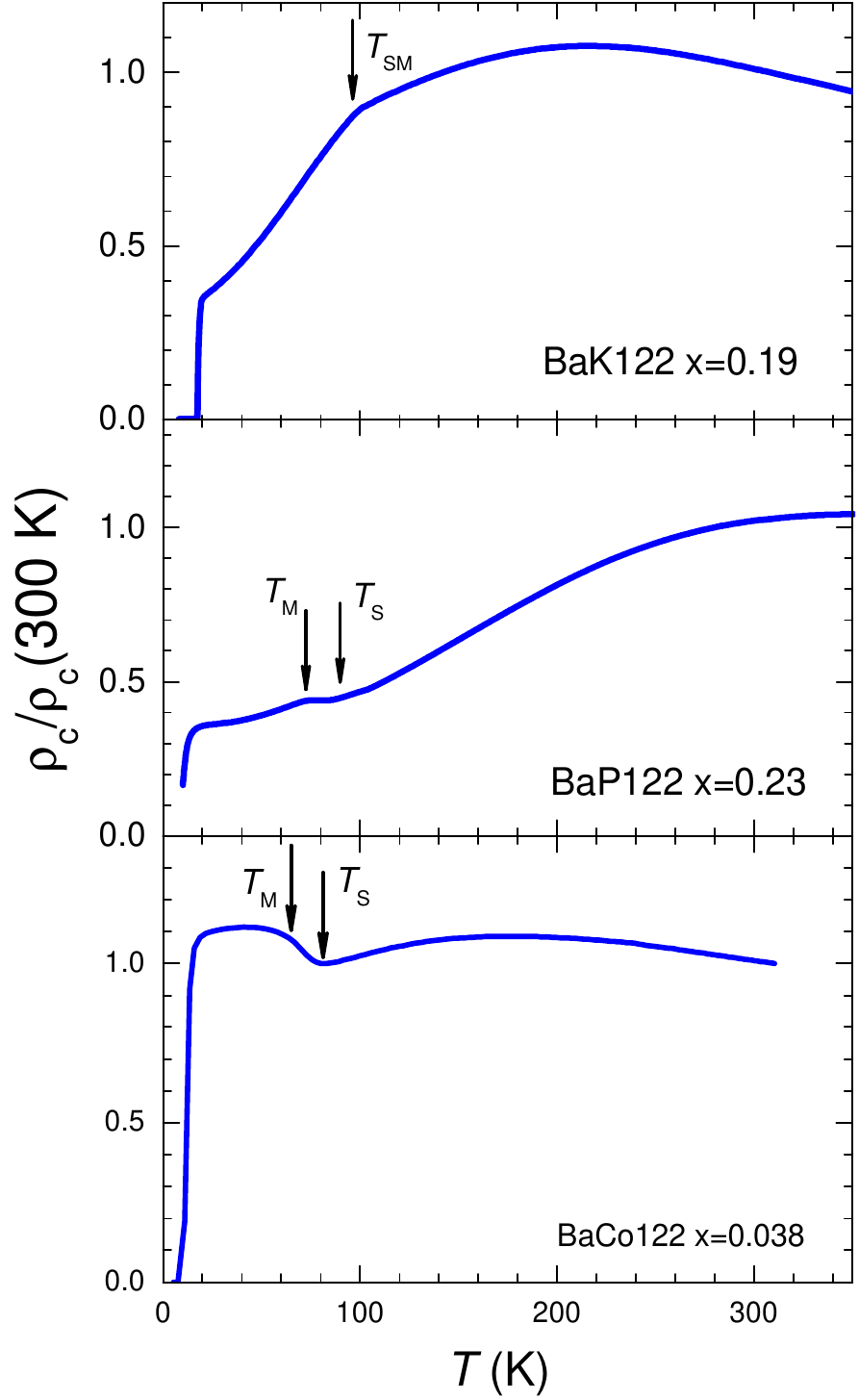}%
\caption{(Color online) Temperature-dependent inter-plane resistivity of underdoped samples of Ba122 based superconductors: (top panel) hole-doped  (Ba$_{1-x}$K$_x$)Fe$_2$As$_2$ $x$=0.19; (middle panel) isoelectron substituted BaFe$_2$(As$_{1-x}$P$_x$)$_2$, $x$=0.23 and (bottom panel) electron-doped Ba(Fe$_{1-x}$Co$_x$)$_2$As$_2$ $x$=0.038.
The resistivity rises significantly below $T_{S}$ in BaCo122, with small slope change at $T_N$, the decrease is smaller in BaP122 and is absent in BaK122. Comparison of the three curves also illustrates different doping evolution of $T_{max}$ for three different types of doping, see phase diagram of the feature in Fig.~\ref{PDpseudogap} below. 
}%
\label{underdoped}%
\end{figure}

Contrary to BaCo122 and BaP122, stripe antiferromagnetic ordering and the tetragonal to orthorhombic structural transition happen simultaneously in BaK122 at a temperature $T_{SM}$=$T_{TO}$=$T_N$ \cite{Avci1,Avci2}. Magnetic ordering reconstructs the Fermi surface, opening a nesting or superzone gaps on electron and hole pockets \cite{MazinSDW}. In hole-doped materials this gap opening, instead of leading to a resistivity increase, leads to an accelerated resistivity decrease (increase of resistivity derivative), suggesting that the main effect comes from a change in the inelastic scattering due to taming down of the contribution of pre-transition  fluctuations of the order parameter. The parts of the Fermi surface which are not affected by the SDW gap \cite{BaKdetwinning,MazinSDW}, enjoy a notably reduced inelastic scattering in the magnetically ordered phase \cite{HallWen,AlloulPRL,detwinning,annealedUchida,BaKdetwinning}. The disorder, inevitably accompanying random distribution of dopant atoms, increases residual resistivity of compounds. This doping disorder is absent in the parent compound, so that decrease of inelastic scattering overcomes the loss of the carrier density and the total conductivity increases below $T_{SM}$. Since the inter-plane transport is dominated by the most warped parts of the Fermi surface \cite{anisotropy}, least affected by the SDW super-zone gap, the inter-plane resistivity should be affected much less by the SDW gap opening than $\rho _a$. This is indeed seen, in BaK122, very similar to BaCo122.

The response of $\rho_c(T)$ to structural/magnetic transition  is distinctly different for hole-doped BaK122, electron-doped BaCo122 and iso-electron substituted BaP122. In Fig.~\ref{underdoped} we compare $\rho_c(T)$ for these different types of doping for compositions with transitions temperatures of order of 100~K, $x$=0.19 in (Ba$_{1-x}$K$_x$)Fe$_2$As$_2$ (top panel), $x$=0.23 in BaFe$_2$(As$_{1-x}$P$_x$)$_2$ (middle panel) and $x$=0.038 in Ba(Fe$_{1-x}$Co$_x$)$_2$As$_2$. The rise of $\rho_c$ in relative units is largest in BaCo122, and smallest (zero) in BaK122, following the same trend as maximum $T_c$ in the series. The comparison of three curves at high temperatures gives a direct hint for different doping evolution of the resistivity crossover maximum at $T_{max}$. 

\subsection{Maximum of inter-plane resistivity at $T_{max}$}

\begin{figure}[tbh]%
\centering
\includegraphics[width=8.5cm]{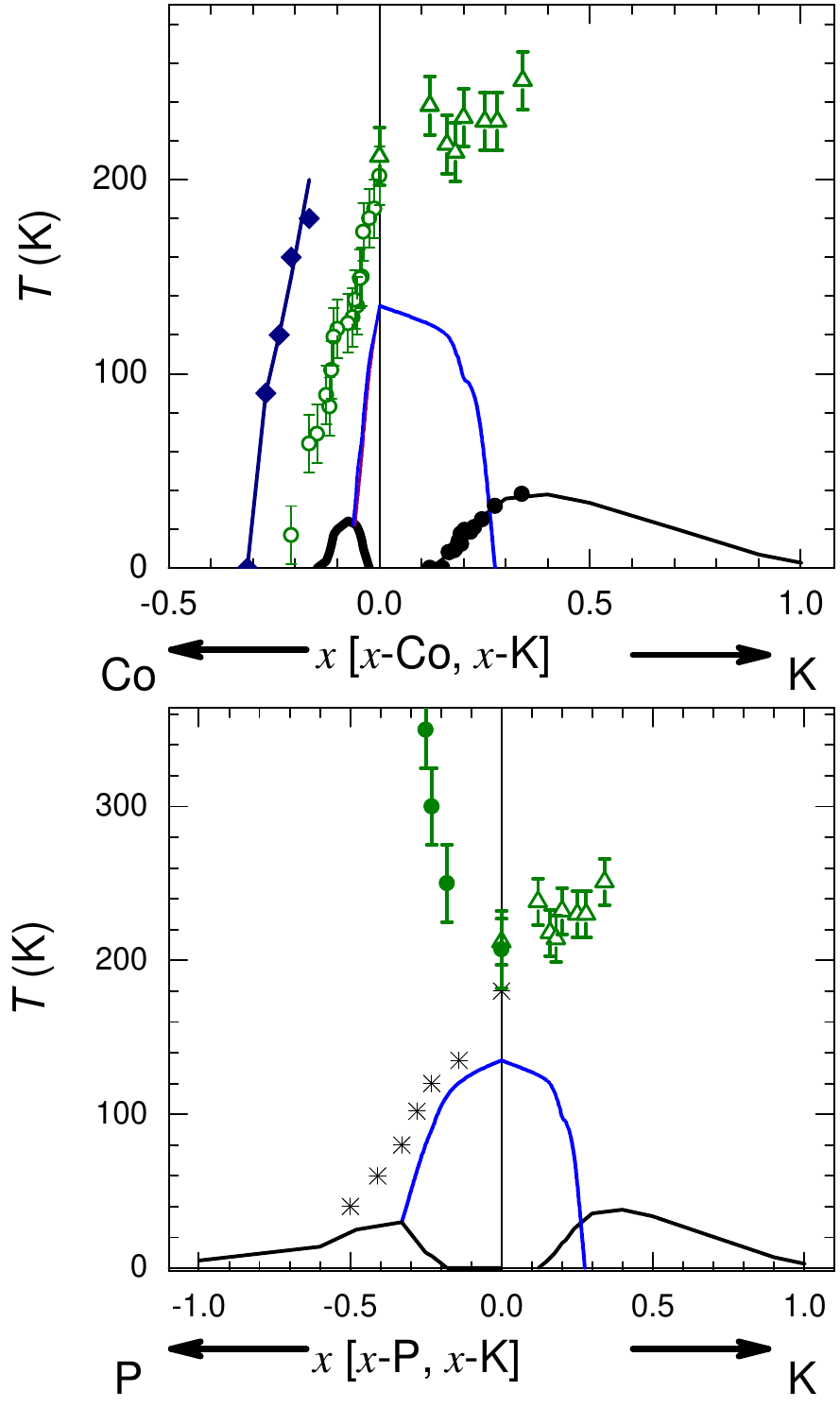}%
\caption{(Color online) Doping phase diagram of hole-doped (Ba$_{1-x}$K$_x$)Fe$_2$As$_2$ shown in comparison with electron-doped Ba(Fe$_{1-x}$Co$_x$)$_2$As$_2$ (top panel) and with iso-electron substituted BaFe$_2$(As$_{1-x}$P$_x$)$_2$ (bottom panel).  Lines show boundaries of orthorhombic (nematic) and magnetically ordered phases and of superconductivity. Open up-triangles show $T_{max}$ in BaK122 as found in this study, open and solid circles show the same feature in BaCo122 and BaP122, diamonds show resistivity minimum found in $\rho_c(T)$ in heavily overdoped BaCo122 \cite{pseudogap}, stars show onset temperature of nematic anomaly in torque measurements in BaP122 \cite{NatureP}. Note asymmetric evolution of the temperature of maximum of the inter-plane resistivity, $T_{max}$, for electron- and hole- doping and much faster increase of the $T_{max}(x)$ in iso-electron substituted BaP122 than in hole-doped BaK122. 
}
\label{PDpseudogap}%
\end{figure}

It is important to notice that the crossover feature at $T_{max}$ in $\rho_c(T)$ is observed through all compositions from parent $x$=0 to 0.34, close to optimal doping. The data of Refs.~\cite{Terashimarhoc,TerashimarhocPRL,ReidPRL,YLiu} show that the crossover feature is observed even in heavily overdoped KFe$_2$As$_2$ ($x$=1). As can be seen in Fig.~\ref{rhooptimal} below, the slope-change feature in the in-plane resistivity $\rho_a(T)$ at around 200~K shows clear correlation with  crossover maximum in the $\rho_c (T)$ at $T_{max}$. This is true for all doping levels, as can be seen from the comparison of Fig.~\ref{rhocdoping} with data of Ref.~\onlinecite{BingShen}.
Both these facts strongly argue against an explanation of the maximum as arising from the balance of several contributions to conductivity. Indeed, the contributions of different sheets of the Fermi surface to the inter-plane transport are determined by their warping, while into in-plane transport by their size. Thus observation of the crossover features at the same temperature on strong variation of the doping level and concomitant changes of the volumes of the electron- and hole- Fermi surface sheets \cite{ARPES,ARPES1} will invoke superficial selection of combinations of carrier densities and mobilities.

In BaCo122 the decrease of the inter-plane resistivity above $T_{max}$ shows a clear correlation with the increase of NMR Knight shift. It also shows clear correlation of the doping range of its existence with the range of $T$-linearly increasing magnetic susceptibility $\chi (T)$. These two observations were the reason for our suggestion of relation of resistivity maximum to onset of carrier activation over a pseudogap. At temperatures below $T_{max}$  both the Knight shift and the inter-plane resistivity in BaCo122 follow the expectations of a metal with the temperature-independent density of states. This density of states becomes temperature-dependent at $T>T_{max}$. Recently similar NMR measurements have been undertaken in optimally doped BaK122 \cite{NMRoptimal}, and found Knight shift which is increasing with temperature and the spin-relaxation rate, 1/$T_1T$, which decreases with temperature and becomes constant above $\sim$200~K. We need to notice though, that the decrease of the inter-plane resistivity, despite being very small, would be very difficult to explain by only a change of the scattering mechanism. It would require activation of carriers by excitations over the partial gap on the Fermi surface (pseudogap).

The largest contributions to the inter-plane transport comes from the most warped sheets of the Fermi surface. According to band structure calculations these are located near the Z point of the Brillouin
zone, on the Fermi surface with dominant contribution of
$d_{3z^2-r^2}$ orbital of iron atom. This band has the weakest nesting and thus should be the least affected by magnetic fluctuations. 
DFT and DMFT calculations suggest that these bands are least renomalized \cite{DFT,DMFT}, and reveal a correlation pseudogap. This orbital selectivity of the correlation pseudogap may explain why the carrier activation is most clearly observed in the inter-plane transport. It does not explain, though, why most localized orbitals affect so strongly in-plane transport only in BaK122 and why the pseudogap value, determining the crossover temperature, is not affected by the doping level change.

\subsection{ Doping evolution of $T_{max}$}

In Fig.~\ref{PDpseudogap} we plot the phase diagram of hole-doped BaK122 in comparison with phase diagrams of electron-doped BaCo122 \cite{pseudogap} and of the iso-electron substituted BaP122 \cite{BaPinterplane}. We focus on the comparison of the salient features of the temperature dependent resistivity, a crossover maximum in $\rho_c (T)$ at $T_{max}$ and a nematic feature found in the in-plane resistivity and torque measurements in BaP122 \cite{NatureP}. First, note electron-hole asymmetry of the $T_{max}(x)$. The $T_{max}$ is rapidly suppressed with electron doping, paving the way to a appearance of a minimum in $\rho_c(T)$ of heavily doped BaCo122 (diamonds in Fig.~\ref{PDpseudogap}). The $T_{max}$ very slightly increases on doping up to the optimal level in BaK122. Second, note that $T_{max}(x)$ dependence in the iso-electron-substituted BaP122 is much stronger than in the hole doped BaK122. This difference leads to a very different temperature-dependent in-plane and inter-plane resistivity at optimal doping.

\subsection {Temperature-dependent resistivity at optimal doping}

\begin{figure}[t]%
\centering
\includegraphics[width=8.5cm]{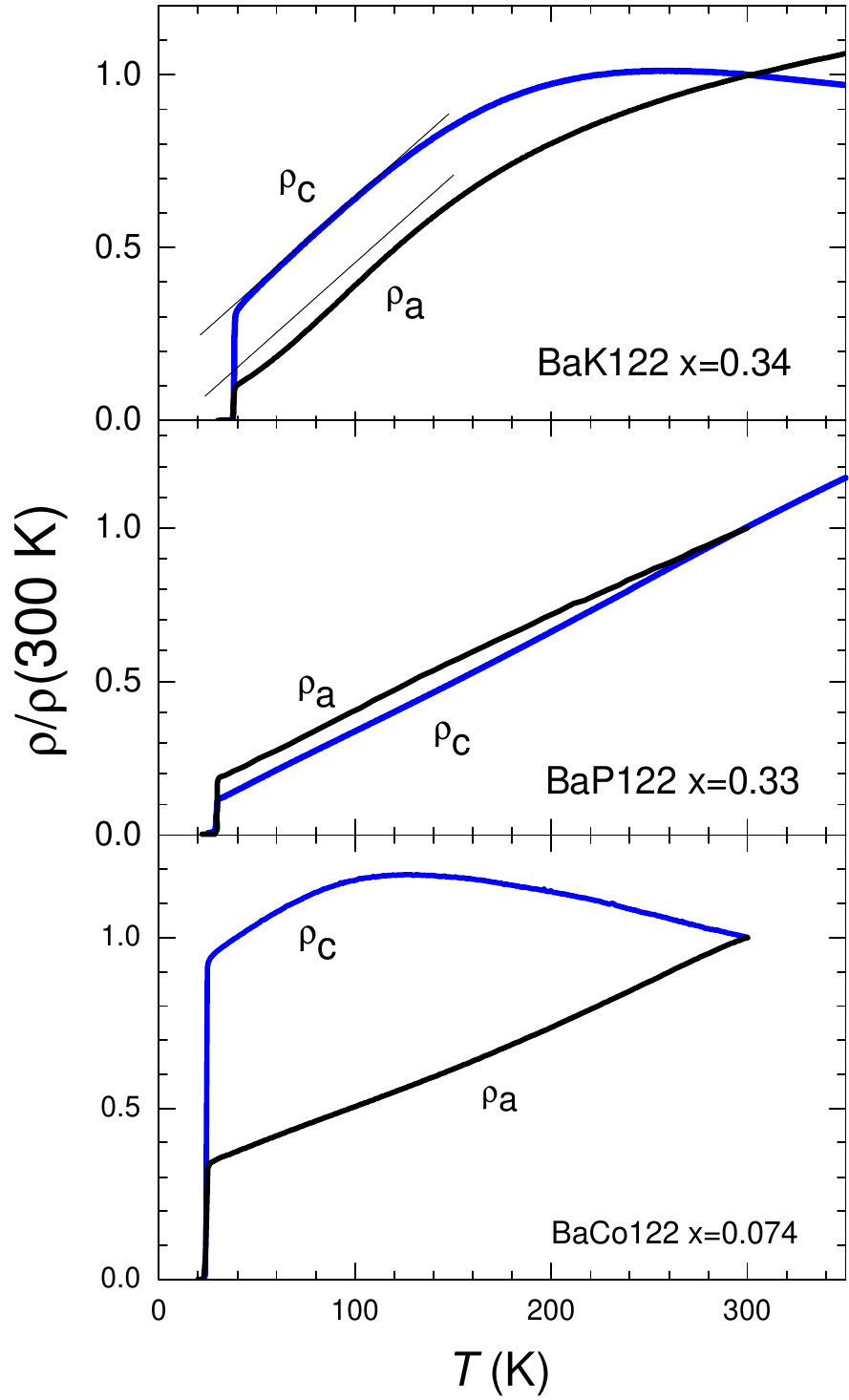}%
\caption{(Color online) Temperature dependent in-plane resistivity (black curves) and inter-plane resistivity (red curves) for samples of hole- doped (Ba$_{1-x}$K$_x$)Fe$_2$As$_2$ (top panel, this study), iso-electron substituted BaFe$_2$(As$_{1-x}$P$_x$)$_2$ (middle panel, Ref.~\onlinecite{BaPinterplane}) and electron-doped Ba(Fe$_{1-x}$Co$_x$)$_2$As$_2$ (bottom panel, Ref.~\onlinecite{pseudogap}). For all curves the data are normalized by the resistivity values at room temperature, $\rho (300K)$.
}%
\label{rhooptimal}%
\end{figure}

In Fig.~\ref{rhooptimal} we show temperature-dependent resistivity of hole-doped BaK122 (top panel), iso-electron-substituted BaP122 (middle panel) and electron-doped BaCo122 (bottom panel), all at optimal doping level. Two features of these curves are prominent. First, in all cases the resistivity is close to $T$-linear above $T_c$. The most clear deviations from linearity are found in $\rho_a(T)$ of BaK122, where some upward curvature can be noticed. This result is similar to previous observations by Bing Shen {\it et al.} \cite{BingShen}. Interestingly, inter-plane resistivity of BaK122 is very close to $T$-linear, which is reminiscent of the anisotropic $T$-linear resistivity at field-tuned quantum critical point of CeCoIn$_5$ \cite{Science}. 
Second, the range of $T$-linear dependence in many cases is confined from above by a crossover temperature $T_{max}$. This is particularly clear for the inter-plane resistivity.

\section{Conclusions}

Measurements of the inter-plane resistivity in BaK122 show that the magnetic/structural transition does not lead to the resistivity increase, i.e., the associated gap does not significantly affect the most warped parts of the Fermi surface, which are important for the inter-plane transport. Upon the suppression of magnetism with doping, the temperature-dependent inter-plane resistivity reveals $T$-linear dependence. This occurs close to the optimal doping just above $T_c$, suggesting the validity of the QCP scenario. This $T-$linear behavior persists up to the pseudogap temperature determined by the maximum in $\rho_c$. Similar to the electron-doped BaCo122 \cite{pseudogap}, we assign the origin of this maximum to the activation of carriers over a pseudogap. This pseudogap is indeed found in some band structure calculations taking into account strong electron correlations \cite{DFT,DMFT}, and is particularly pronounced in the Fermi surface parts with larger contributions from $d_{3z^2-r^2}$ and $d_{x^2-y^2}$ orbitals. The former is also responsible for the most warped $\zeta$ sheet of the KFe$_2$As$_2$ Fermi surface \cite{TerashimaFermisurface,ARPESFS}. This orbital selectivity of the pseudogap may provide natural explanation as to why the pseudogap feature is affecting mostly the inter-plane transport.

Despite the effect of doping in multi-band metallic system may be quite complicated, comparison of the hole-doped BaK122 with the electron-doped BaCo122 shows significant difference. A pseudogap resistive crossover at $T_{max}$ in the inter-plane resistivity vanishes with doping in BaCo122 but remains intact in BaK122. The crossover affects temperature-dependent in-plane resistivity in BaK122, however, it does not in BaCo122. The pseudogap crossover temperature in BaK122 increases much slower than in the iso-electron substituted BaP122.

Finally we would like to point to a certain similarity in the critical behavior of the inter-plane resistivity in BaK122 and in CeCoIn$_5$. In CeCoIn$_5$, a true critical behavior at a field-tuned QCP \cite{PaglioneQCP,BianchiQCP} with $T$-linear resistivity and violation of the Wiedemann-Franz law is observed for transport along the tetragonal $c$-axis \cite{Science}, while transport along the plane obeys the Wiedemann-Franz law \cite{nonvanishing}. This is similar to the difference in the temperature dependence of $\rho_a(T)$ and $\rho_c(T)$ in BaK122, with the latter being more linear at optimal doping.

\section{Acknowledgements}

This work was supported by the U.S. Department of Energy (DOE), Office of Science, Basic Energy Sciences, Materials Science and Engineering Division. The research was performed at the Ames Laboratory, which is operated for the U.S. DOE by Iowa State University under contract DE-AC02-07CH11358. Work in China was supported by the Ministry of Science and Technology of China, project 2011CBA00102. Part of the work done in the University of Sherbrooke was supported by the Canadian Institute for Advanced Research and a Canada Research Chair, and it was funded by NSERC, FQRNT, and CFI.


\end{document}